\documentclass[conference]{IEEEtran}
\IEEEoverridecommandlockouts
\usepackage{cite}
\usepackage{amsmath,amssymb,amsfonts}
\usepackage{algorithmic}
\usepackage{graphicx}
\usepackage{textcomp}
\usepackage{xcolor}
\usepackage{balance}
\usepackage{hyperref}
\def\BibTeX{{\rm B\kern-.05em{\sc i\kern-.025em b}\kern-.08em
    T\kern-.1667em\lower.7ex\hbox{E}\kern-.125emX}}
\begin{document}

\title{Enhancing Multivariate Time Series-based Solar Flare Prediction with Multifaceted Preprocessing and Contrastive Learning}

\author{
    \IEEEauthorblockN{MohammadReza EskandariNasab, Shah Muhammad Hamdi, Soukaina Filali Boubrahimi}
    \IEEEauthorblockA{
        \textit{Department of Computer Science}, \\
        \textit{Utah State University}, \\
        Logan, UT 84322, USA \\
        \{reza.eskandarinasab, s.hamdi, soukaina.boubrahimi\}@usu.edu
    }
}

\maketitle

\begin{abstract}
Accurate solar flare prediction is crucial due to the significant risks that intense solar flares pose to astronauts, space equipment, and satellite communication systems. Our research enhances solar flare prediction by utilizing advanced data preprocessing and classification methods on a multivariate time series-based dataset of photospheric magnetic field parameters. First, our study employs a novel preprocessing pipeline that includes missing value imputation, normalization, balanced sampling, near decision boundary sample removal, and feature selection to significantly boost prediction accuracy. Second, we integrate contrastive learning with a GRU regression model to develop a novel classifier, termed ContReg, which employs dual learning methodologies, thereby further enhancing prediction performance. To validate the effectiveness of our preprocessing pipeline, we compare and demonstrate the performance gain of each step, and to demonstrate the efficacy of the ContReg classifier, we compare its performance to that of sequence-based deep learning architectures, machine learning models, and findings from previous studies. Our results illustrate exceptional True Skill Statistic (TSS) scores, surpassing previous methods and highlighting the critical role of precise data preprocessing and classifier development in time series-based solar flare prediction.
\end{abstract}


\section{Introduction}

Solar flares pose a significant threat to humans and equipment in space due to the intense radiation they emit \cite{hazard}. These flares can lead to rapid and significant surges in radiation levels, encompassing extreme-ultraviolet, X-rays, and gamma-rays across the electromagnetic spectrum. They are classified logarithmically based on their peak soft X-ray flux, with categories A, B, C, M, and X increasing in intensity, starting from a flux of \(10^{-8} \, \text{W/m}^2\). Consequently, the peak soft X-ray flux of an X-class flare is typically ten times more intense than that of an M-class flare and one hundred times more intense than that of a C-class flare.

Recent studies on flare prediction have emphasized the use of data science techniques, specifically utilizing spatiotemporal magnetic field data from the Helioseismic Magnetic Imager (HMI) \cite{hmi, hmi2, hmi3} on the Solar Dynamics Observatory (SDO) \cite{sdo, ahmadzadeh}. This data is transformed into multivariate time series (MVTS) instances to predict flares within specific temporal windows \cite{swansf}. Each instance includes 24 photospheric magnetic field parameters represented as time series. These MVTS instances are categorized into five classes: flare-quiet (FQ) instances (including both flare-quiet and A-class) and flare classes of increasing intensity (B, C, M, X). An essential dataset for this research is the \emph{Space Weather Analytics for Solar Flares (SWAN-SF)} \cite{swansf}, which is created from solar photospheric vector magnetograms by the Spaceweather HMI Active Region Patch (SHARP) series \cite{sharp}.

Data collected to address real-world problems is seldom clean or immediately usable, even with thorough screening processes. Such datasets often come with challenges related to the subject matter or the data collection method. These challenges, which include missing values, multi-scaled attributes, class overlap, class imbalance, and irrelevant features \cite{velanki2024}, are common in many nonlinear dynamical systems such as streamflow prediction \cite{streamflow}, neuro-developmental disorder prediction \cite{hamdineuro}, and auditory attention detection \cite{EskandariNasab2024}. These significant challenges underscore the necessity of developing advanced data preprocessing and machine learning-based approaches for solar flare prediction.

Our research focuses on developing an innovative data preprocessing pipeline and a novel classification method to substantially improve the performance of solar flare classification. Our contributions are as follows:

\begin{enumerate}
    
    \item We introduce a multifaceted preprocessing pipeline to address the challenges associated with the SWAN-SF dataset. This pipeline includes several key stages: a missing value imputation technique that combines next-value and previous-value imputation, global z-score normalization, and balanced sampling through SMOTE \cite{smote} and random under sampling (RUS) to address class imbalance. Additionally, a ‘near decision boundary sample removal’ (NDBSR) \cite{ndbsr} technique is employed to eliminate border samples, thereby enhancing the classifier’s performance. Furthermore, using a GRU model, we evaluate the impact of each photospheric magnetic field parameter to underscore the importance of feature selection. This multifaceted preprocessing substantially improves the classification accuracy of solar flares.
    
    \item Moreover, we present a novel classification technique called \emph{ContReg (Contrastive and Regression-based learning)} to further enhance flare prediction. This technique integrates two types of learning: a GRU-based contrastive learning network \cite{contrastive} with triplet loss and a GRU-based regression network. Additionally, we introduce a new integrated loss function, providing an innovative approach to classifying solar flares. This method utilizes both the flare category (binary) and the peak soft X-ray flux for classification, resulting in superior performance.

\end{enumerate}

The paper further discusses related work in Section \ref{sec:relatedwork}, followed by an explanation of the methodologies in Section \ref{sec:method}. It then presents the experiments and results in Section \ref{sec:experiments}, and finally concludes the paper in Section \ref{sec:conclusion}.

\section{Related Work}
\label{sec:relatedwork}

Previous studies on solar flare prediction have focused on the development and optimization of machine learning algorithms to enhance prediction accuracy \cite{Nishizuka2017, Nishizuka2018}. Among methods based on photospheric vector magnetograms, the study by Bobra and Couvidat (2015) \cite{b2} utilized preflare instantaneous values of active region magnetic field parameters to predict solar flares using a support vector machine (SVM) classifier. The paper by Hamdi et al. (2017) \cite{b3} proposed a flare prediction technique by extracting time series samples of active region parameters and applying k-Nearest Neighbors (k-NN) classification on the univariate time series. The study by Ahmadzadeh et al. (2021) \cite{ahmadzadeh} tackled specific challenges in solar flare forecasting, such as class imbalance and temporal coherence, discussing strategies such as under-sampling and over-sampling to manage class imbalance in the SWAN-SF dataset and highlighting the importance of proper data splitting and validation techniques to ensure model robustness against temporal coherence. The study by Muzaheed et al. (2021) \cite{b5} employed Long Short-Term Memory (LSTM) networks for effective end-to-end classification of MVTS in solar flare prediction, outperforming traditional models and demonstrating the potential of deep learning. Similarly, the study by Hamdi et al. (2022) \cite{b6} developed a novel approach combining Graph Convolutional Networks (GCNs) with LSTM networks, effectively capturing both spatial and temporal relationships in solar flare prediction and surpassing other baseline methods. The paper by Alshammari et al. (2022) \cite{b7} addressed the forecasting of magnetic field parameters related to flaring events using a deep sequence-to-sequence learning model with batch normalization and LSTM networks. In our recent study \cite{eskandari_apjs}, we presented a novel preprocessing pipeline that leverages advanced techniques such as FPCKNN imputation, LSBZM normalization \cite{notebooks}, and a multi-stage sampling approach. This effort resulted in a fully preprocessed version of the SWAN-SF dataset, which is now publicly available \cite{clean_swansf}. By providing this preprocessed version, we aimed to streamline future research on solar flare prediction, enabling researchers to bypass the time-consuming preprocessing stage and focus directly on model development and analysis.

While previous studies have each focused on specific aspects of preprocessing, in this paper, we present a simplistic yet comprehensive and novel data preprocessing pipeline with multiple stages designed to address all preprocessing challenges of the SWAN-SF dataset. Additionally, we introduce an advanced method for classifying solar flares by utilizing both the binary labels of flares and their peak soft X-ray flux.

\section{Methodology}
\label{sec:method}

\begin{figure}
\centerline{\includegraphics[width=0.5\textwidth]{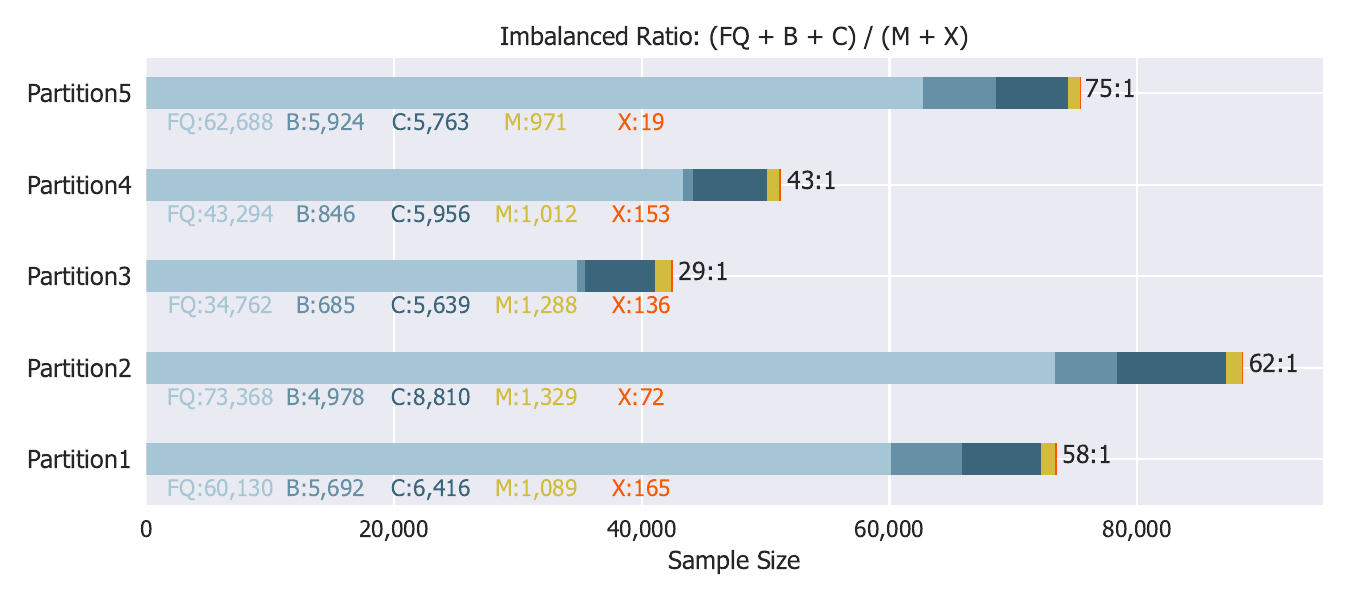}}
\caption{The figure presents a stacked bar chart illustrating the distribution of different solar flare classes within each partition of the SWAN-SF dataset. This visualization is based on the current methodology of time series slicing used in SWAN-SF, which involves steps of 1 hour, an observation period of 12 hours, and a prediction span of 24 hours. Each slice of the MVTS is categorized according to the most intense flare reported within its prediction window.}
\label{fig:swansf}
\end{figure}

Fig. \ref{fig:swansf} illustrates the class distribution within partitions of the SWAN-SF dataset. Each MVTS record in SWAN-SF includes 24 time series of magnetic field parameters from an active region (AR), captured at 12-minute intervals over 12 hours, resulting in 60 time steps. Our research concentrates on a binary classification task using this MVTS data to distinguish between major-flaring (X and M classes) and minor-flaring (C, B, and FQ classes) ARs. Initially, we discuss the preprocessing stages that we have implemented and subsequently provide a detailed explanation of the ContReg classifier.

\subsection{Preprocessing Pipeline}
Despite the SWAN-SF dataset presenting numerous data quality challenges \cite{ahmadzadeh}, we thoroughly investigate these issues to preprocess the dataset effectively and achieve high classification performance, as indicated by TSS (True Skill Statistic) score. The TSS is a metric used to evaluate imbalanced binary classification models, defined as the difference between recall and the false positive rate. Our preprocessing methodology comprises several steps designed to address these issues sequentially, enhancing the TSS score with each step. The following sections outline these steps in detail.

\begin{figure}
  \centering
  \includegraphics[width=0.5\textwidth]{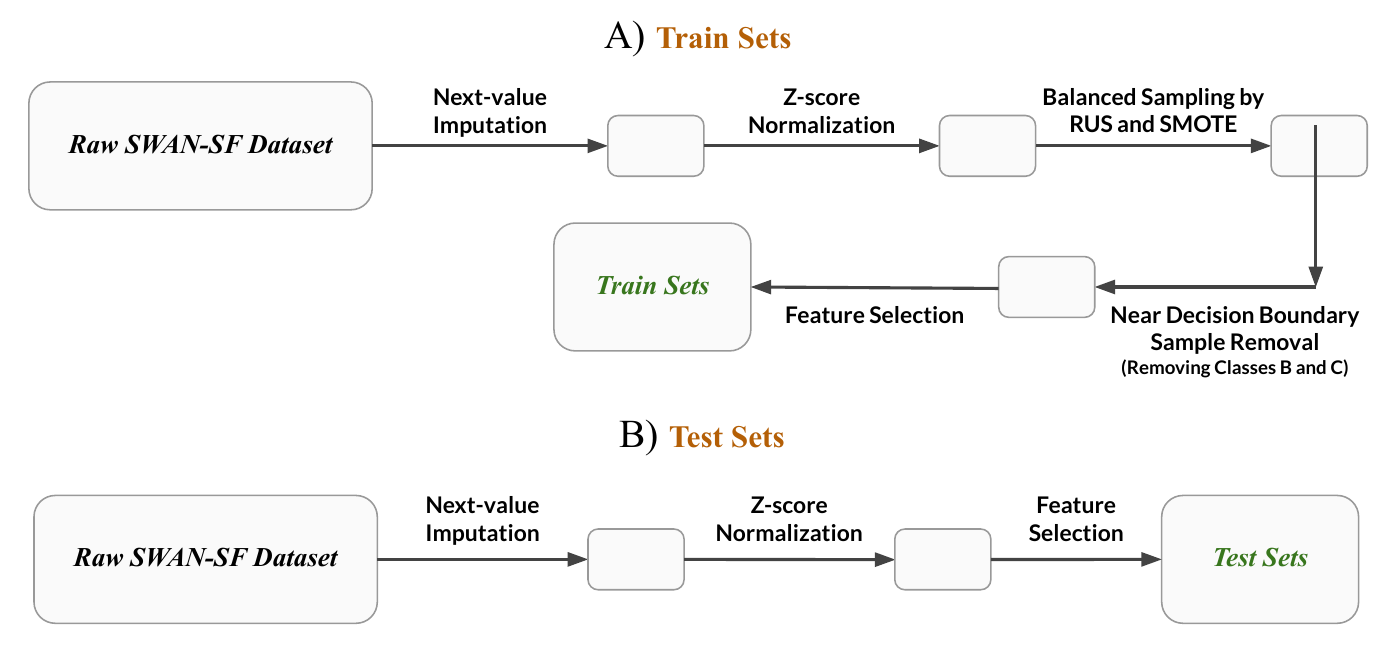}
  \caption{The figure demonstrates the different stages of our preprocessing pipeline for training and testing sets. Panel A showcases the training sets, while Panel B showcases the testing sets. No sampling methodologies are applied to the testing sets to avoid biased results.}
  \label{fig:preprocessing}
\end{figure}

\subsubsection{\textbf{Imputation}}
In the case of SWAN-SF, there are two types of missing values. In some instances, only certain timestamps are missing, while in others, entire timestamps for a feature are missing. Specifically, the occurrence of missing entire timestamps primarily affects FQ class samples. Given the substantial number of FQ class samples in each partition, we exclude them from the dataset. However, when only some timestamps are missing, imputation becomes essential, particularly for important classes such as X-class samples, where the number of samples is limited, necessitating their retention. To impute these samples, we employ a combination of next-value and previous-value imputation. Initially, we aim to impute a missing timestamp with the next available one. If no subsequent timestamp is available, we use the previous available timestamp. Next-value and previous-value imputation maintain the temporal consistency of the data, preserving inherent patterns and trends \cite{imputation2}. These methods are particularly suitable for time series data, where the missing values are likely similar to their neighboring values.

\subsubsection{\textbf{Normalization}}
Normalization ensures that each attribute of a dataset equally influences the prediction, thereby enhancing the accuracy and efficiency of the models \cite{ahmadzadeh}. The SWAN-SF dataset exhibits unique and heterogeneous ranges for each attribute, necessitating the use of normalization. Z-score normalization standardizes the values of a dataset so that they have a mean of 0 and a standard deviation of 1. We employ global Z-score normalization, which utilizes the dataset’s overall statistics, adjusting based on the mean and standard deviation values of each feature rather than each sample individually. Therefore, for each feature, the time series data from all samples are first concatenated. Subsequently, z-score normalization is performed on the concatenated series. The normalized values are then used in place of the original values in the dataset.

\subsubsection{\textbf{Balanced Sampling}}
An imbalanced dataset, as shown in Fig. \ref{fig:swansf}, can substantially diminish the accuracy of a classification model by causing a bias towards the majority class, thereby leading to frequent misclassification of the minority class \cite{adasyn}. Consequently, the model’s ability to make accurate predictions for minority class instances is significantly hindered. To address this issue, it is essential to generate additional synthetic samples for the underrepresented minority class using over-sampling techniques. However, if the number of majority class samples is disproportionately large, it is also necessary to decrease their quantity through under-sampling techniques. Failing to do so would require generating an excessive amount of synthetic data for the minority class to achieve balance, which could result in overfitting.

\begin{figure}
  \centering
  \includegraphics[width=0.45\textwidth]{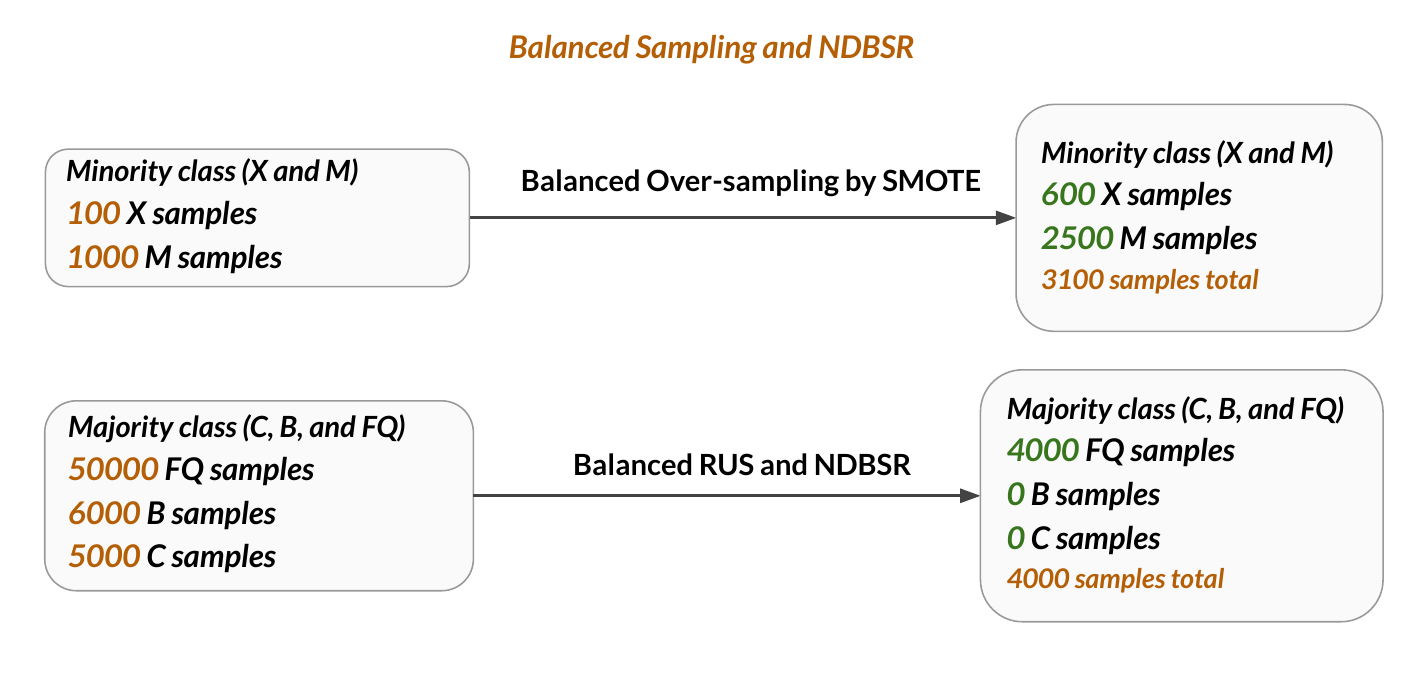}
  \caption{The figure illustrates the concept of Balanced Sampling, which includes both Balanced Over-Sampling and Balanced Random Under Sampling (RUS), along with the NDBSR strategy, a more sophisticated approach to addressing class imbalance. In the provided example, representing an approximation of the first partition from the SWAN-SF dataset, synthetic samples are generated for subclasses X and M in a controlled manner. The aim is to avoid generating an excessive number of synthetic samples, thereby preventing them from dominating the original samples. At the same time, a higher proportion of synthetic samples is generated for subclass X (500\%) compared to subclass M (150\%) to ensure a balanced representation between subclasses. Furthermore, in alignment with the NDBSR strategy, samples from subclasses B and C are completely removed from the minor-flaring class (comprising classes FQ, B, and C), while only a small portion of samples from class FQ is retained and utilized. This approach ensures that the distribution of the minor-flaring class (FQ) is aligned with that of the major-flaring samples (X and M classes), thereby promoting a balanced representation across the classes.}
  \label{fig:balancedsampling}
\end{figure}

In the case of the SWAN-SF dataset, each class, including major-flaring, contains subclasses, such as X and M flares, which also exhibit imbalance. Therefore, we developed a balanced sampling approach that employs over-sampling (via SMOTE) and under-sampling (via RUS) techniques to ensure balance not only between the major-flaring and minor-flaring classes but also among the subclasses within each class. Furthermore, it is important to note that even with advanced deep learning-based data augmentation techniques, the quality of the generated samples does not match that of real samples \cite{timegan}. Thus, we should avoid generating an excessive number of synthetic samples compared to real ones, as this could distort the model towards the synthetic data. Our balanced sampling technique, therefore, ensures that a reasonable number of synthetic samples is generated, preventing them from dominating the real samples. An example of our balanced sampling technique can be seen in Fig. \ref{fig:balancedsampling}.

\subsubsection{\textbf{NDBSR}}
Methods such as NDBSR \cite{ndbsr} can be employed to improve the model’s ability to distinguish between overlapping classes. NDBSR involves identifying and removing data points situated too close to the decision boundary. The SWAN-SF dataset presents significant challenges for classification due to extensive overlap between classes. This complexity arises from the dataset’s nature: flares are categorized into five classes based on the intensity of peak soft X-ray flux, resulting in samples near the class boundaries being very similar to each other.

In the realm of solar flare prediction, accurately predicting all major-flaring events (class X and M) is crucial due to the severe consequences of missing an X-class flare \cite{xclassimpact}. Therefore, it is recommended to achieve a high recall even if it results in a low rate of false positives, rather than aiming for a low recall with no false positives. To address the issue of class overlap, we retain all samples from the major-flaring classes (X and M) while removing all samples from the B and C classes. Consequently, the minor-flaring class includes only FQ class samples. This approach significantly reduces the overlap between the major-flaring and minor-flaring classes, ensuring a higher recall score for the major-flaring events. As a result, this improvement in recall directly contributes to an enhanced TSS score, reflecting better overall model performance and more accurate flare classification.

\begin{table*}
\centering
\caption{Selected active region magnetic field parameters in SWAN-SF dataset}
\label{table:attributes}
\small
\begin{tabular}{|l|c|c|}
\hline
\textbf{Abbreviation} & \textbf{Description} & \textbf{Formula} \\
\hline
ABSNJZH \cite{leka2003photospheric} & Absolute value of the net current helicity & $H_{c_{a b s}} \propto\left|\sum B_{z} \cdot J_{z}\right|$  \\
\hline
SAVNCPP \cite{leka2003photospheric} & Sum of the modulus of the net current per polarity & $J_{z_{sum}} \propto \vert \sum^{B_z^+} J_zdA \vert  + \vert \sum^{B_z^-} J_zdA \vert$ \\
\hline
TOTBSQ \cite{fisher2012global} & Total magnitude of Lorentz force & $F \propto \sum B^{2}$\\
\hline
TOTPOT \cite{leka2003photospheric} & Total photospheric magnetic free energy density & $\rho_{tot} \propto \sum (\boldsymbol{B}^{Obs} - \boldsymbol{B}^{Pot})^2 dA$\\
\hline
TOTUSJH \cite{leka2003photospheric} & Total unsigned current helicity & $H_{c_{\text {total }}} \propto \sum B_{z} \cdot J_{z}$\\
\hline
TOTUSJZ \cite{leka2003photospheric} & Total unsigned vertical current & $J_{z_{total}}= \sum\left|J_{z}\right| d A$\\
\hline
\end{tabular}
\scriptsize
\end{table*}

\subsubsection{\textbf{Feature Selection}}
The SWAN-SF dataset consists of 24 photospheric magnetic field features, which have been used collectively in most previous studies to classify solar flares. However, according to the study by Alshammari et al. (2024) \cite{Alshammari_2024}, many of these 24 features not only fail to contribute to the prediction and classification of flares but also introduce noise and degrade the performance of the classifier. Therefore, as shown in Table \ref{table:attributes}, we select our own set of features, consisting of only six out of the original 24 features. To select these six features, we employ a GRU classifier, providing each feature separately as input data to obtain its TSS score in the binary classification of flares. We then select the top features that achieve the highest TSS scores. The features listed in Table \ref{table:attributes} attain the highest individual TSS scores, while the remaining features perform significantly worse in comparison. This decision significantly improves the TSS score of the preprocessing pipeline and also reduces the training time considerably, as we use only 25 percent of the original features.

\subsection{ContReg Classifier}
Previous research on predicting solar flares has largely depended on using binary labels to classify flares into major-flaring and minor-flaring categories \cite{b2}. However, this approach can be significantly enhanced by incorporating the peak soft X-ray flux of each sample, as these provide a quantitative measure of flare intensity. Accordingly, the ContReg classifier incorporates both flares labels and X-ray flux intensity to improve its predictive performance. This is conceptually similar to how the AAD-GCQL model \cite{EskandariNasab2024} combines GRU and CNN architectures to leverage both spatial and temporal features in EEG signals, improving the detection of auditory attention.

However, traditional sequence-to-sequence deep learning models such as LSTM, GRU, and RNN often struggle to distinguish between similar patterns in major- and minor-flaring events, frequently failing to recognize subtle distinctions, which leads to inadequate classification performance \cite{contrastive}. Therefore, the ContReg utilizes contrastive learning to map the input into a lower-dimensional space, where it learns to differentiate between the two categories of flares (major- and minor-flaring) by focusing on the subtle differences in their features through the use of triplet loss.

\subsubsection{\textbf{Architecture}}
\begin{figure}
  \centering
  \includegraphics[width=0.5\textwidth]{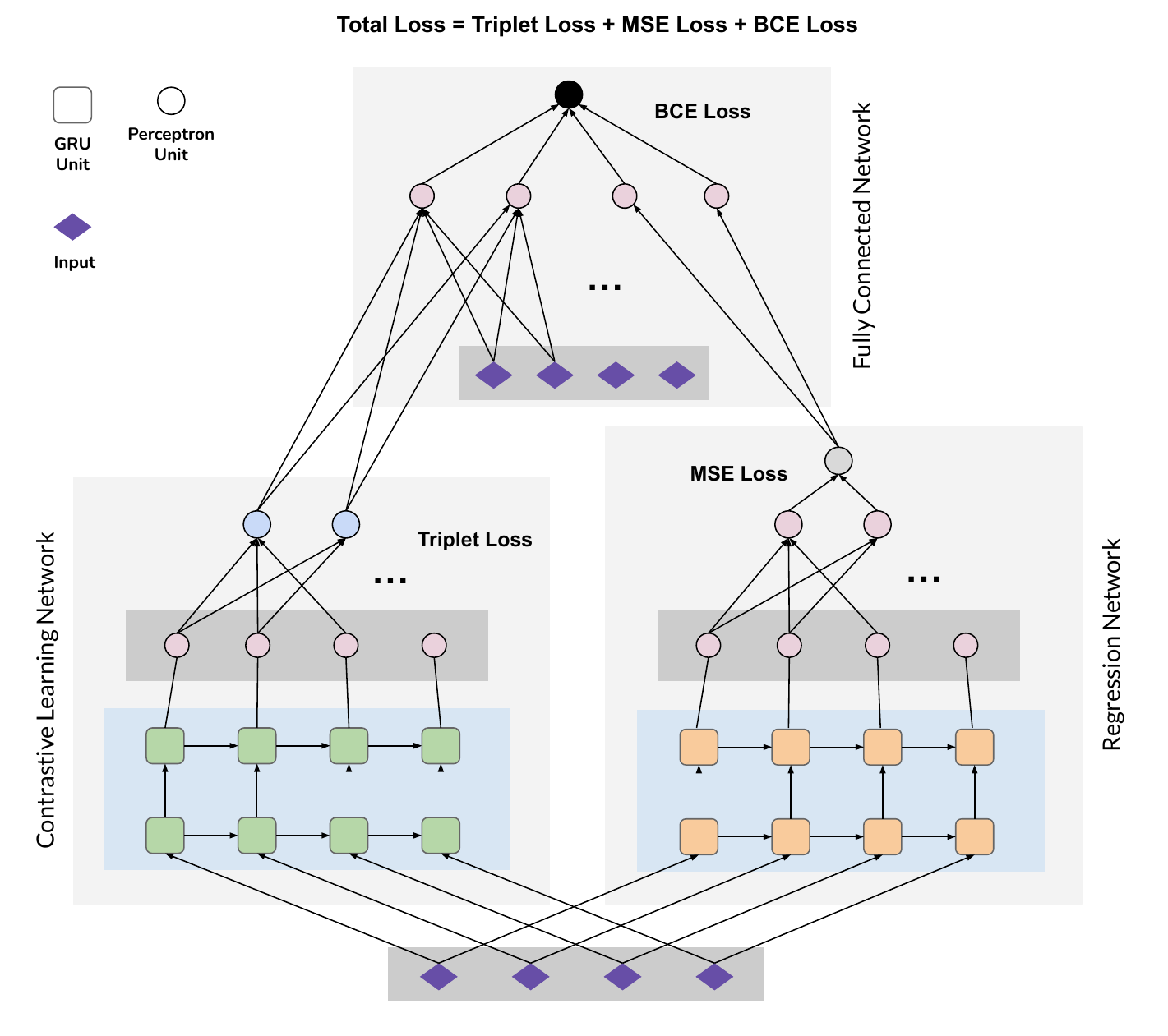}
  \caption{The figure illustrates the architecture of ContReg, which employs three individual networks and utilizes a combined loss function to train the network and classify solar flare events. The three dots in the figure illustrate the concept of a fully connected layer. The technique combines contrastive learning with regression to create concise information that is fed into the final fully connected neural network, along with the actual input, to achieve higher classification performance.
}
  \label{fig:arch}
\end{figure}

As shown in Fig. \ref{fig:arch}, ContReg consists of three networks, with the first being a GRU-based contrastive learning network. This network is designed to output data in a lower-dimensional space, ensuring that the representations for the two categories of flares are sufficiently distinct. This distinction facilitates the subsequent classifier’s ability to accurately distinguish between these categories. Contrastive learning is a technique that aims to minimize the distance between similar data points while maximizing the distance between dissimilar ones \cite{contrastive}. It uses a triplet loss function to ensure that an anchor sample is closer to a positive sample (same category) than to a negative sample (different category). Second, ContReg includes a GRU-based regression network to output the peak soft X-ray flux of the flare, providing the final classifier with additional information. Third, it incorporates a fully connected neural network as the final classifier where the inputs are the outputs of the contrastive learning network, the regression model, and the original input data. The output of this network is the binary label of the flare. This approach ensures that the final classifier benefits from both the distinct feature representations learned through contrastive learning and the quantitative flare intensity provided by the regression model, resulting in a more robust flare classification.

\subsubsection{\textbf{Triplet and Total Loss}}

The triplet loss of our contrastive learning network (\(L_{\text{Triplet}}\)) employs cosine similarity to measure the differences between the anchor and positive samples, as well as between the anchor and negative samples. Cosine similarity measures the cosine of the angle between two vectors, with values ranging from -1 to 1, where 1 indicates identical vectors and -1 indicates opposite vectors. Specifically, the triplet loss function ensures that an anchor sample is closer to a positive sample (same class) than to a negative sample (different class) by a margin \(\alpha\). The cosine similarity and the triplet loss are defined as follows:

\begin{equation}
\cos(a, b) = \frac{a \cdot b}{\|a\| \|b\|}
\end{equation}

\begin{equation}
L_{\text{Triplet}} = \max(0, \cos(a, n) - \cos(a, p) + \alpha)
\end{equation}

where \(a\) is the anchor sample, \(p\) is the positive sample, \(n\) is the negative sample, and \(\alpha\) is a margin to ensure a significant difference between positive and negative pairs.

Consequently, this loss function facilitates the network in learning a representation where samples from the same class are grouped together, and samples from different classes are well-separated. To further enhance the robustness of the network, for each anchor, we select four negative and four positive samples from the batch. We then calculate the differences between the anchor and the four positive samples, as well as the anchor and the four negative samples, and use the average of these differences to compute the loss. This approach enhances the overall training of the model by providing a more reliable learning signal.

Meanwhile, the loss function of the regression network is the mean square error (MSE), which is calculated as follows, where \(y_i\) is the true value and \(\hat{y}_i\) is the predicted value:

\begin{equation}
L_{\text{MSE}} = \frac{1}{n} \sum_{i=1}^{n} (y_i - \hat{y}_i)^2
\end{equation}

Additionally, the loss function of the fully connected network is the binary cross entropy (BCE). This loss function evaluates the performance of a classification model whose output is a probability value between 0 and 1. The BCE is defined as follows, where \(y_i\) is the true binary label and \(\hat{y}_i\) is the predicted probability:

\begin{equation}
L_{\text{BCE}} = -\frac{1}{n} \sum_{i=1}^{n} [y_i \log(\hat{y}_i) + (1 - y_i) \log(1 - \hat{y}_i)]
\end{equation}

Finally, the ContReg model is trained as a single network, where its loss function, referred to as the total loss, is a combination of these three losses. The total loss (\(L_{\text{Total}}\)) is defined as follows, where \(\lambda_1\), \(\lambda_2\), and \(\lambda_3\) are weighting factors that balance the contributions of the triplet loss, MSE loss, and BCE loss, respectively:

\begin{equation}
L_{\text{Total}} = \lambda_1 L_{\text{Triplet}} + \lambda_2 L_{\text{MSE}} + \lambda_3 L_{\text{BCE}}
\end{equation}

\section{Experiments}
\label{sec:experiments}

The Python repository for our preprocessing pipeline and ContReg classifier, along with detailed documentation of the implementation and hyperparameters used, is publicly available for extensive review and application \footnote{The codebase for this paper is accessible here: \href{https://github.com/samresume/PreprocessingPipeline-ContRegClassifier-SWANSF}{https://github.com/samresume/PreprocessingPipeline-ContRegClassifier-SWANSF}}.

\begin{figure}
  \centering
  \includegraphics[width=0.50\textwidth]{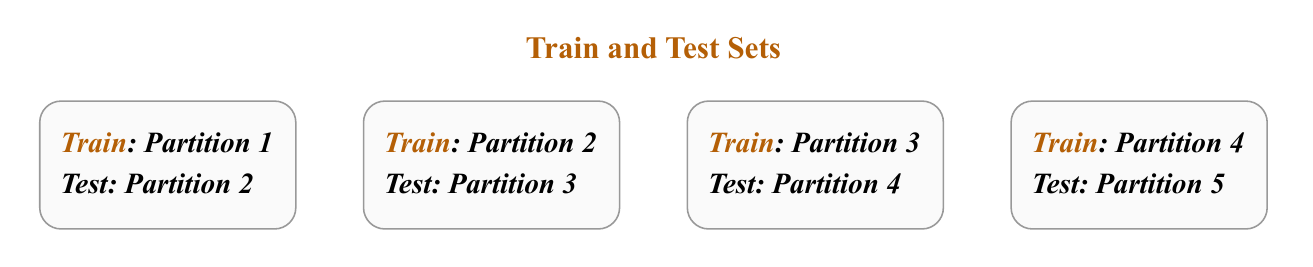}
  \caption{This figure showcases four distinct train-test sets employed in each classification experiment. This approach ensures a more comprehensive and accurate assessment of algorithms across all dataset partitions. In real-world time series forecasting, it is recommended that the training set precedes the test set chronologically, as the goal is always to predict the future. This approach leads to a more accurate and meaningful evaluation.}
  \label{fig:traintest}
\end{figure}

For our experiments, we utilize four unique train-test combinations of the SWAN-SF dataset, as illustrated in Fig. \ref{fig:traintest}. Given the temporal ordering of the partitions, it is optimal to select combinations that are consecutive. This approach ensures that the training set precedes the test set in terms of the temporal sequence, which is a crucial factor in our analysis.

\subsection{Evaluation Metrics}

\begin{table*}[ht]
\centering
\caption{Performance gain achieved by each step of the preprocessing pipeline}
\label{table:preprocessing}
\small
\begin{tabular}{|l|c|c|c|}
\hline
\textbf{Experiment} & \textbf{Mean TSS ± STD} & \textbf{Mean Recall ± STD} & \textbf{Mean HSS ± STD} \\
\hline
Only I (LSTM) & 0.173 ± 0.06 & 0.452 ± 0.29 & 0.051 ± 0.02 \\
\hline
I-N (LSTM) & 0.655 ± 0.06 & 0.856 ± 0.05 & 0.190 ± 0.07 \\
\hline
I-N-Smote (LSTM) & 0.611 ± 0.06 & 0.798 ± 0.08 & 0.183 ± 0.04 \\
\hline
I-N-BS (LSTM) & 0.639 ± 0.07 & 0.858 ± 0.13 & 0.182 ± 0.09 \\
\hline
I-N-BS-NDBSR (LSTM) & 0.690 ± 0.05 & 0.894 ± 0.03 & \textbf{0.192 ± 0.05} \\
\hline
\textbf{I-N-BS-NDBSR-FS (LSTM)} & \textbf{0.801 ± 0.03} & \textbf{0.945 ± 0.01} & 0.185 ± 0.04 \\
\hline
Ahmadzadeh 2021 (SVM) & 0.613 ± 0.04 & NA & NA \\
\hline
Alshammari 2024 (Transformer) & 0.693 ± 0.04 & NA & NA \\
\hline
\end{tabular}
\end{table*}

\begin{table*}[ht]
\centering
\caption{Comparison of our classifier with baseline methods and previous studies}
\label{table:classifiers}
\small
\begin{tabular}{|l|c|c|c|}
\hline
\textbf{Experiment} & \textbf{Mean TSS ± STD} & \textbf{Mean Recall ± STD} & \textbf{Mean HSS ± STD} \\
\hline
k-NN & 0.699 ± 0.05 & 0.816 ± 0.04 & 0.193 ± 0.05 \\
\hline
RandomForest & 0.700 ± 0.07 & 0.781 ± 0.07 & \textbf{0.247 ± 0.05} \\
\hline
SVM & 0.782 ± 0.04 & 0.908 ± 0.05 & 0.203 ± 0.06 \\
\hline
1D-CNN & 0.762 ± 0.05 & 0.893 ± 0.06 & 0.197 ± 0.07 \\
\hline
RNN & 0.807 ± 0.02 & 0.943 ± 0.03 & 0.195 ± 0.05 \\
\hline
LSTM & 0.801 ± 0.03 & 0.945 ± 0.01 & 0.185 ± 0.04 \\
\hline
GRU & 0.805 ± 0.03 & 0.925 ± 0.03 & 0.215 ± 0.05 \\
\hline
\textbf{ContReg (Ours)} & \textbf{0.846 ± 0.01} & \textbf{0.975 ± 0.02} & 0.213 ± 0.03 \\
\hline
Ahmadzadeh 2021 (SVM) & 0.613 ± 0.04 & NA & NA \\
\hline
Alshammari 2024 (Transformer) & 0.693 ± 0.04 & NA & NA \\
\hline
\end{tabular}
\end{table*}

We employ the TSS as the evaluation metric \cite{Bloomfield_2012} since accuracy is inadequate for imbalanced datasets and can produce misleadingly high scores. The TSS is a valuable metric for evaluating imbalanced datasets, particularly in solar flare prediction. It effectively balances the model's recall and its ability to limit the false positive rate, thus providing a comprehensive measure of model performance. The optimal value of the TSS is +1, indicating perfect performance with a recall of 1 and a false positive rate of 0. The TSS is calculated as:
\begin{equation}
    \textbf{TSS} = \text{Recall} - \text{False Positive Rate} = \frac{\text{TP}}{\text{TP} + \text{FN}} - \frac{\text{FP}}{\text{FP} + \text{TN}}
\end{equation}

Additionally, recall is utilized as the secondary evaluation metric due to the critical importance of the model’s capability to predict all major-flaring events, even at the expense of incurring some False Positives as errors. Furthur, another metric employed in previous studies is the Heidke Skill Score (HSS) \cite{Mason_2010}, as provided by the Space Weather Prediction Center. The optimal value of the HSS is +1, indicating perfect performance. The HSS is calculated as:

\begin{equation}
\textbf{HSS} = \frac{2 \times [(\text{TP} \times \text{TN}) - (\text{FN} \times \text{FP})]}{(\text{TP} + \text{FN}) \times (\text{FN} + \text{TN}) + (\text{TP} + \text{FP}) \times (\text{FP} + \text{TN})}
\end{equation}

\subsection{Baseline Techniques}

We compare our multifaceted preprocessing and ContReg classifier with both baseline techniques and previous studies to demonstrate their superior performance in predicting solar flares. In particular, we reference Ahmadzadeh et al. (2021) \cite{ahmadzadeh} and Alshammari et al. (2024) \cite{Alshammari_2024}, which have reported the highest TSS for binary classification of solar flares using the SWAN-SF dataset.

\subsubsection{Preprocessing Baselines}

To illustrate the impact of each stage in our preprocessing pipeline, we analyze the performance improvements achieved at each step by incrementally adding the components. We begin with Imputation (I), then introduce global Z-score Normalization (I-N), followed by Balanced Sampling (I-N-BS). Next, we apply the NDBSR technique (I-N-BS-NDBSR), and finally, we incorporate Feature Selection to complete the pipeline (I-N-BS-NDBSR-FS). Additionally, we compare the Balanced Sampling stage (I-N-BS) with the use of SMOTE for over-sampling (I-N-Smote) to demonstrate the effectiveness of our approach.

\subsubsection{Classification Baselines}
We compare the ContReg classifier with various sequence-to-sequence deep learning classifiers, including LSTM, RNN, GRU, and 1D-CNN. Additionally, we evaluate our technique against classical machine learning methods, including SVM, Random Forest, and k-NN, to achieve a comprehensive evaluation.

\subsection{Results and Discussion}

According to Table \ref{table:preprocessing}, our complete preprocessing pipeline achieves a mean TSS of 0.801 and a mean recall of 0.945, outperforming the results of Ahmadzadeh et al. (2021) and Alshammari et al. (2024), who reported mean TSS scores of 0.613 and 0.693, respectively. This performance is based on the four train-test combinations discussed earlier. As each stage was added to the pipeline, we observed incremental improvements in TSS. Initially, with only imputation (I), the mean TSS was 0.173. By incorporating global z-score normalization (I-N), this improved to 0.655. The addition of balanced sampling and NDBSR techniques (I-N-BS-NDBSR) further raised the mean TSS to 0.690. Finally, after applying feature selection (I-N-BS-NDBSR-FS), the mean TSS reached 0.801, demonstrating the effectiveness of the full pipeline. The recall score also consistently improved with each successive step in the preprocessing pipeline. Additionally, as shown in Table \ref{table:classifiers}, the integration of our ContReg classifier with the preprocessing pipeline results in a mean TSS of 0.846 and a mean recall of 0.975, demonstrating superior performance compared to the previous studies. Deep learning-based techniques, including GRU, LSTM, and RNN, achieved better results than traditional machine learning models such as SVM, k-NN, and Random Forest. The “NA” values in the table indicate that these values were not provided by the previous studies.

\begin{figure}
  \centering
  \includegraphics[width=0.5\textwidth]{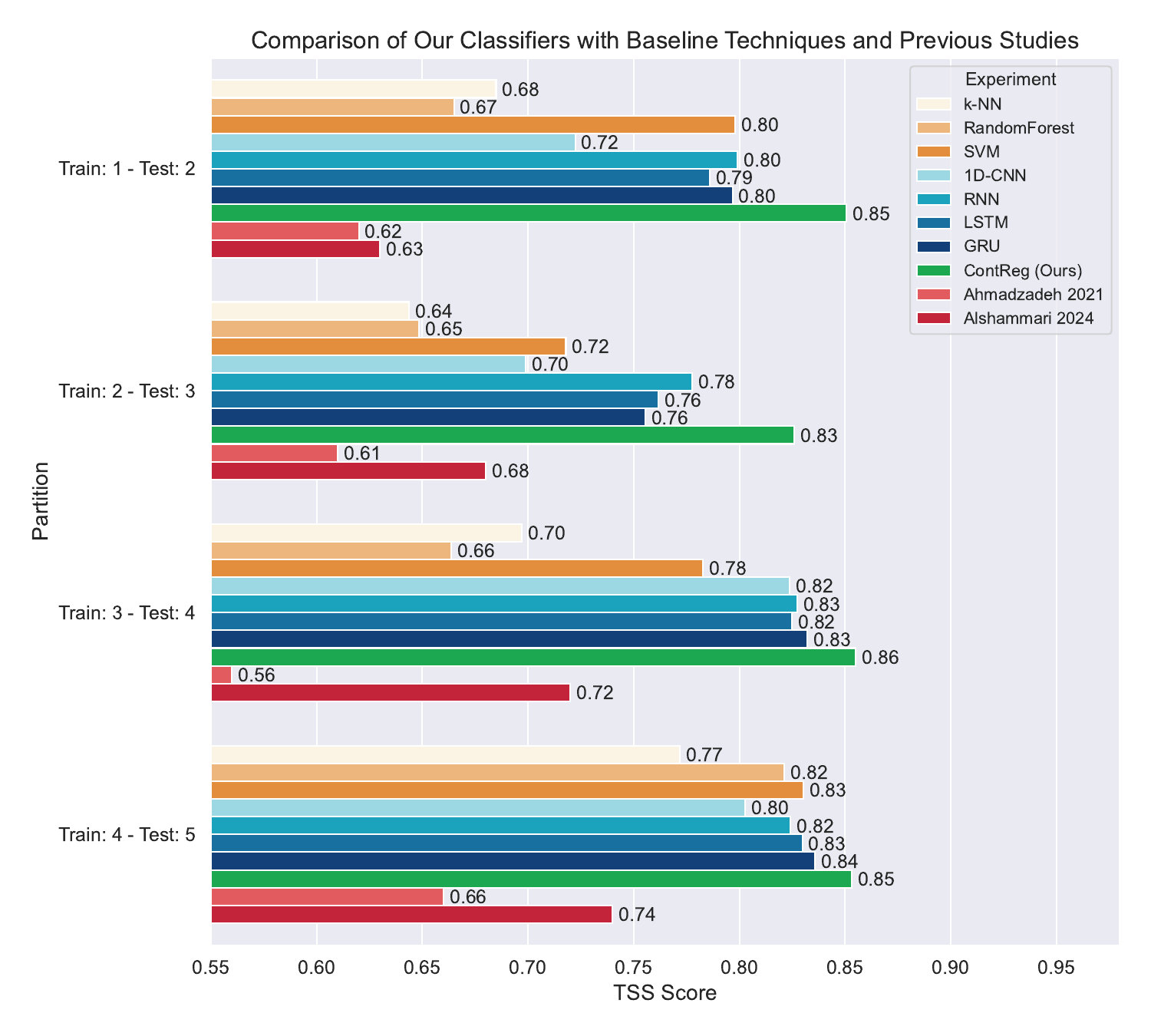}
  \caption{Comparison of our classifier (ContReg) with baseline techniques and previous studies. In these experiments, our complete preprocessing pipeline is applied to both our classifier (ContReg) and the baseline techniques.}
  \label{fig:classifiers}
\end{figure}

\begin{figure}
  \centering
  \includegraphics[width=0.5\textwidth]{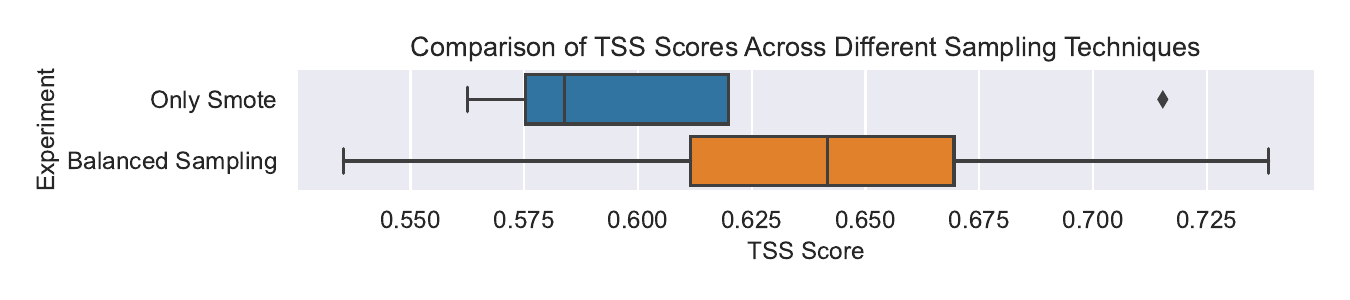}
  \caption{Comparison of Balanced Sampling and Over-Sampling (SMOTE only) on various training and testing combinations of the SWAN-SF dataset using different classifiers.}
  \label{fig:bs_smote}
\end{figure}

However, evaluating the impact of the preprocessing pipeline and the ContReg classifier on each train-test combination is essential to assess the consistency of the technique in improving performance. As illustrated in Fig. \ref{fig:preprocessing}, our comprehensive preprocessing, when combined with an LSTM classifier, consistently outperforms the methodologies presented in the studies by Ahmadzadeh et al. (2021) and Alshammari et al. (2024) across all train-test combinations. Referring to both Figs. \ref{fig:preprocessing} and \ref{fig:bs_smote}, the application of our balanced sampling technique significantly improves the TSS score compared to the sole use of SMOTE for over-sampling, which often leads to overfitting. Additionally, normalization substantially enhances the TSS score compared to using only imputation. Moreover, the NDBSR technique further boosts the TSS score, highlighting the importance of addressing class overlap issues. Fig. \ref{fig:classifiers} demonstrates that the ContReg classifier, combined with our preprocessing pipeline, not only significantly outperforms previous studies but also surpasses commonly used deep learning and machine learning techniques, consistently across all train-test combinations.

\begin{figure}
\centering
\includegraphics[width=0.5\textwidth]{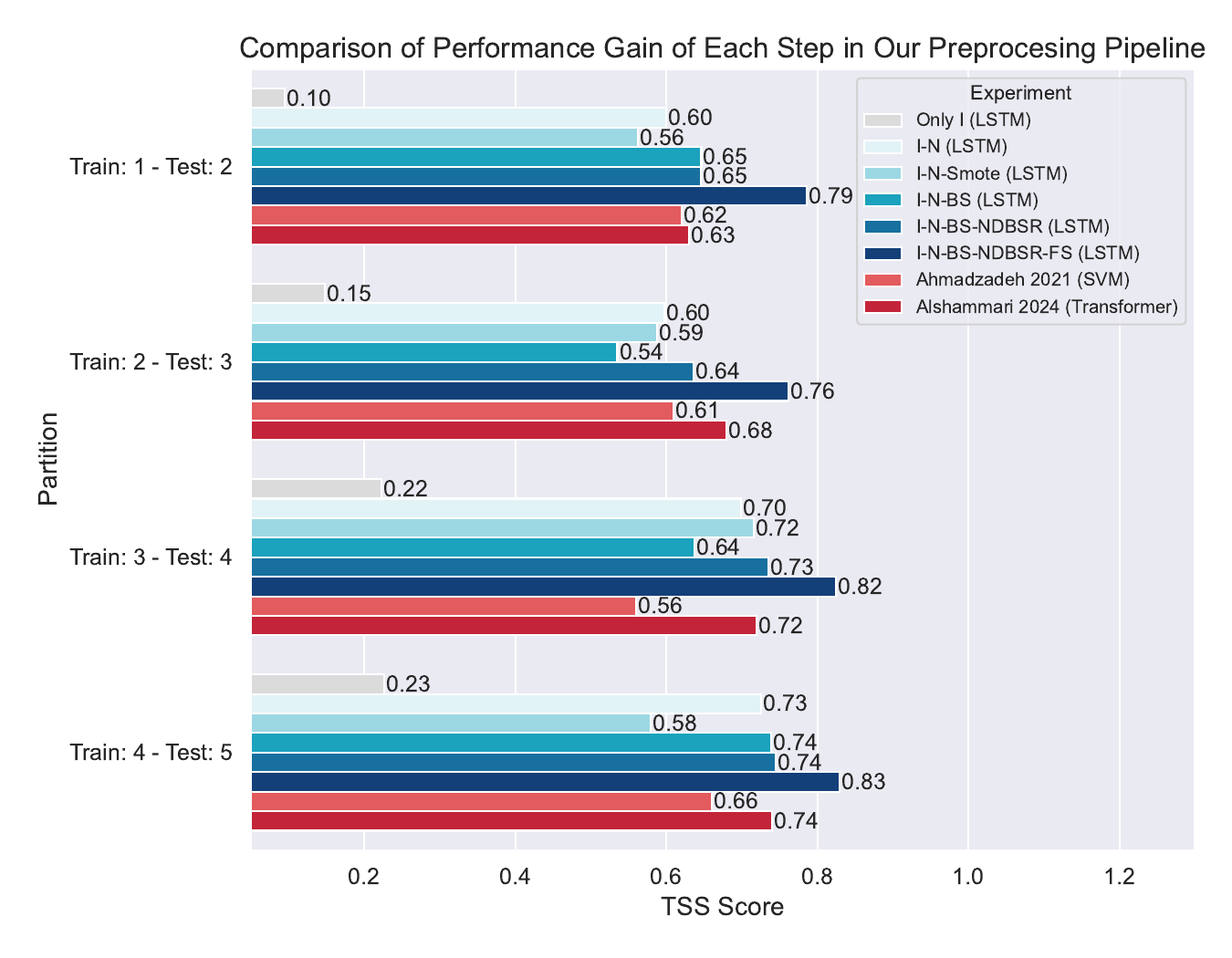}
\caption{Comparison of performance gain at each stage of our preprocessing pipeline, along with comparison to previous studies. A simple LSTM classifier was employed to demonstrate the performance gain at each stage of our preprocessing pipeline and to minimize the classifier’s impact on the results.}
\label{fig:preprocessing}
\end{figure}

\section{Conclusion and future Work}
\label{sec:conclusion}

Through extensive experiments incorporating our multifaceted preprocessing approach and the ContReg classifier, we have demonstrated a significant improvement in both TSS and recall scores across all training and testing combinations of the SWAN-SF dataset’s partitions. A mean TSS of 0.801 was achieved by incorporating our multifaceted preprocessing and using only a simple LSTM, highlighting the critical role of a precise preprocessing pipeline in solar flare prediction. This is particularly relevant when dealing with challenging datasets such as SWAN-SF, which pose distinct preprocessing challenges. Furthermore, by incorporating our preprocessing pipeline with our ContReg classifier, we have further improved the TSS score to a mean of 0.846, which significantly outperforms all previous studies and makes solar flare prediction a more robust task. For future research, we plan to develop a robust time series generation technique utilizing Adversarial Autoencoders. This approach aims to create a more accurate over-sampling method specifically designed for MVTS data.

\section{Acknowledgment}

This project has been supported by the Division of Atmospheric and Geospace Sciences within the Directorate for Geosciences through NSF awards \#2301397, \#2204363, and \#2240022, as well as by the Office of Advanced Cyberinfrastructure within the Directorate for Computer and Information Science and Engineering under NSF award \#2305781.

\balance

\end{document}